\documentclass[twocolumn,aps]{revtex4}
\usepackage{epsfig}
\usepackage{psfig}
\usepackage{amsmath}

\newcommand{\be}{\begin{equation}}   
\newcommand{\ee}{\end{equation}}   
\newcommand{\bea}{\begin{eqnarray}}   
\newcommand{\eea}{\end{eqnarray}}

\begin{document}
\title{The axis of evil}
\author{Kate Land and Jo\~{a}o Magueijo }
\address{Theoretical Physics Group, Imperial College, Prince Consort Road, 
London SW7 2BZ, UK}

\begin{abstract}
We examine previous claims for a preferred axis at $(b,l)\approx (60,-100)$
in the cosmic radiation anisotropy,
by generalizing the concept of multipole planarity
to any shape preference (a concept we define mathematically). 
Contrary to earlier claims, we find that the amount of power
concentrated in planar modes for $\ell=2,3$ is not inconsistent
with isotropy and Gaussianity.
The multipoles' alignment, however, 
is indeed anomalous, and extends up to $\ell=5$ rejecting statistical 
isotropy with a probability in excess of 99.9\%. 
There is also an uncanny correlation of azimuthal phases
between $\ell=3$ and $\ell=5$. 
We are unable to blame these effects on foreground contamination
or large-scale systematic errors. We show how this reappraisal
may be crucial in identifying the theoretical model behind the anomaly.
\end{abstract}

\pacs{PACS Numbers: *** }
\date{Feb 11, 2005}

\maketitle

The homogeneity and isotropy of the Universe -- also known as
the Copernican principle -- is a major postulate of modern cosmology.
Obviously this assumption does not imply 
exact homogeneity and isotropy, but merely that the 
observed cosmological inhomogeneities are random 
fluctuations around a uniform background, extracted from a homogeneous and 
isotropic statistical ensemble. One may expect that the ever
improving observations of CMB fluctuations should 
lead to the greatest vindication
of this principle. Yet, there have been a number of disturbing
claims of evidence for a preferred direction in the 
Universe~\cite{teg,copilet,ral,erik1,hbg,us,erik2,erik3,hansen,vielva},
making use of the state of the art 
WMAP first year results~\cite{wmap}. These claims
have potentially very damaging 
implications for the standard model of cosmology. 

It has been suggested that a preferred direction in CMB fluctuations
may signal a non-trivial cosmic topology (e.g.~\cite{riaz,teg,dodec,sperg}), a
matter currently far from settled. The preferred axis could 
also be the result of anisotropic expansion,
possibly due to strings, walls or magnetic fields~\cite{arj},
or even the result of an intrinsically inhomogeneous
Universe~\cite{moffatinh}.
Such claims remain controversial; more mundanely
the observed ``axis of evil'' could be the result of galactic foreground
contamination or large scale unsubtracted systematics 
(see~\cite{fmg,band,inter,jooes} for past examples).
A closer inspection of the emergence of this preferred axis is
at any rate imperative. 

In addressing the issue of statistical isotropy 
it is of paramount importance to work in harmonic space, a matter that
has not always been fully appreciated.  Consider
a full-sky map, $\frac{\Delta T}{T}(\hat{\mathbf r})$.
It is well known that the space of such functions is {\it not}
an irreducible representation of the rotation group, i.e. it
contains invariant subspaces, the so-called harmonic multipoles. 
One should therefore perform an expansion into Spherical Harmonic 
functions (eigenfunctions of $L^2$ and $L_z$, where ${\mathbf L}$
is the angular momentum operator)
\begin{eqnarray}
\frac{\Delta T}{T}(\hat{\mathbf r})=\sum_\ell \delta T_\ell=
\sum_{\ell m}a_{\ell m}Y_{\ell m}(\hat{\mathbf r})
\label{almdef}
\end{eqnarray}
and examine the problem multipole by multipole.
Each multipole $\ell$ is an irreducible representation 
of SO(3) and a systematic examination of statistical isotropy can 
now be carried out.

The close relation between statistical anisotropy and non-Gaussianity
(loosely the emergence of structured shapes)
has been extensively discussed in the literature~\cite{conf,fermag}.
For an isotropic Gaussian process each $a_{\ell m}$ is an
independent Gaussian random variable characterised by variance
$C_\ell$. In each realization the power $C_\ell$ is
randomly, uniformly distributed among all $m$s, and therefore
there is no evident ``shape'' to the multipole. One finds conflict
with this prediction if, for a given multipole $\ell$, one identifies an axis 
${\mathbf n}_\ell$ for which an uncanny amount of power is concentrated 
in a given $m$-mode, once the $z$-axis is aligned with ${\mathbf n}$.
In general reorienting the axis destroys this feature, so
the onset of $m$-preference is also a measure of statistical
anisotropy. 

For example planar configurations (see~\cite{teg,copi,copilet})
have a $z$-axis' orientation where all the power is concentrated in 
the $m=\pm\ell$ modes. They maximally break cylindrical symmetry around 
the $z$-axis. On the contrary, cylindrically symmetric 
configurations (``Jupiters'') have a $z-$axis orientation where
all the power is in the $m=0$ mode, so that the sphere is divided
into latitude bands, or zones, without any longitude variation in the
temperature. Even in these extreme cases if we reorient the axis
the power spreads to other $m$-modes and the $m$-preference is
obscured. So the issue of $m$ preference is closely linked with 
that of anisotropy.

\begin{figure}
\centerline{\psfig{file=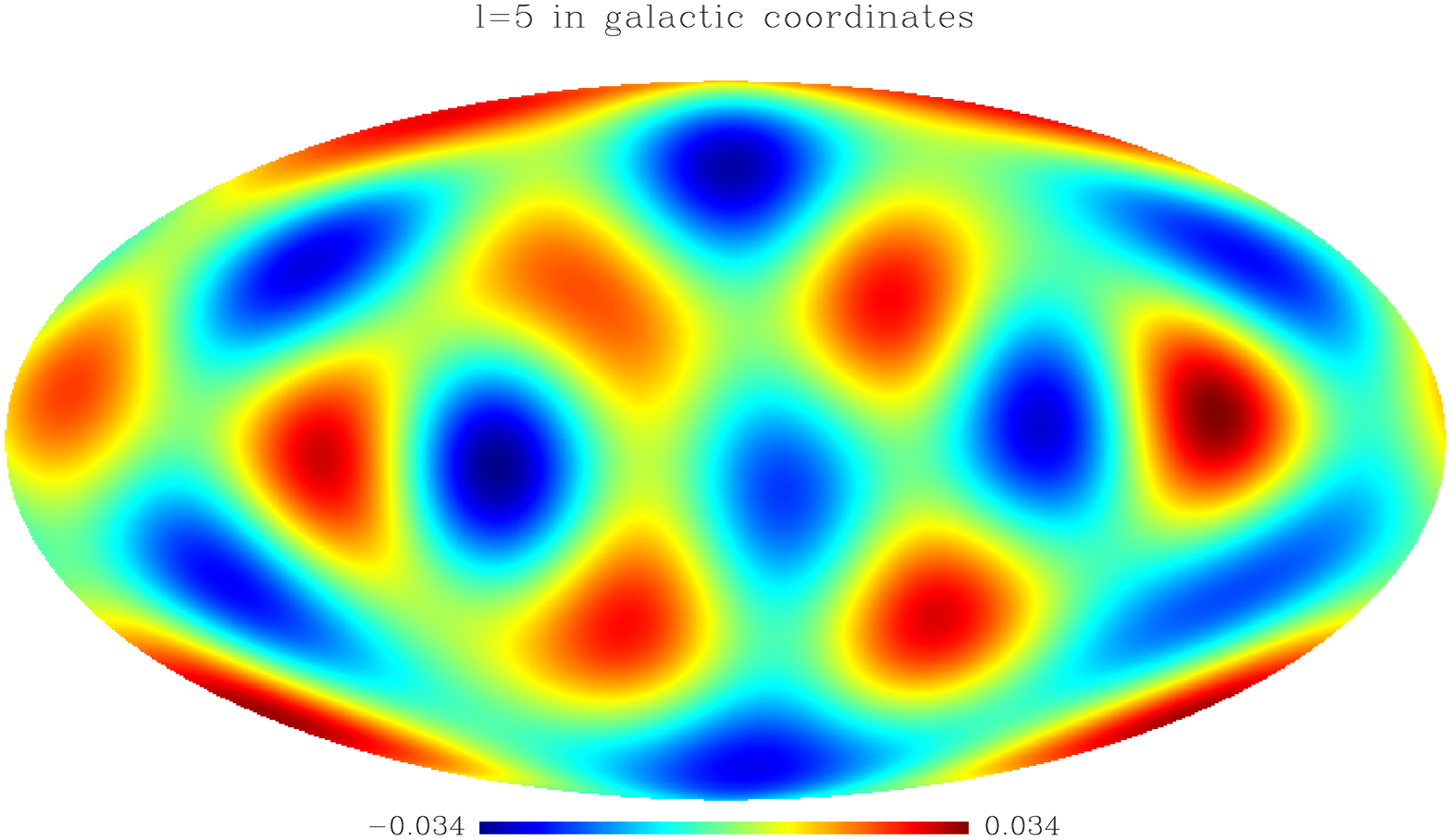,angle=90,width=9cm}}
\centerline{\psfig{file=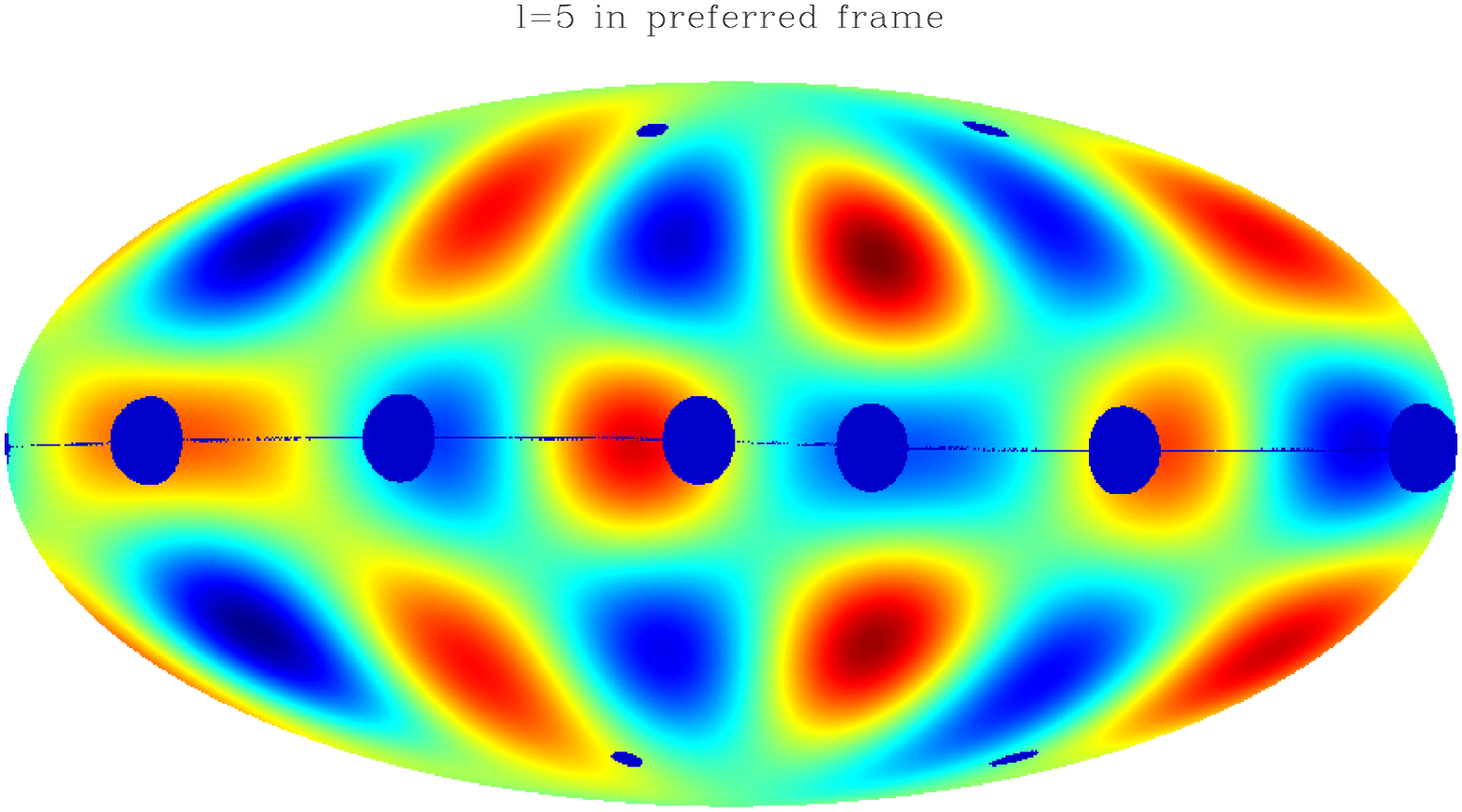,angle=90,width=9cm}}
\centerline{\psfig{file=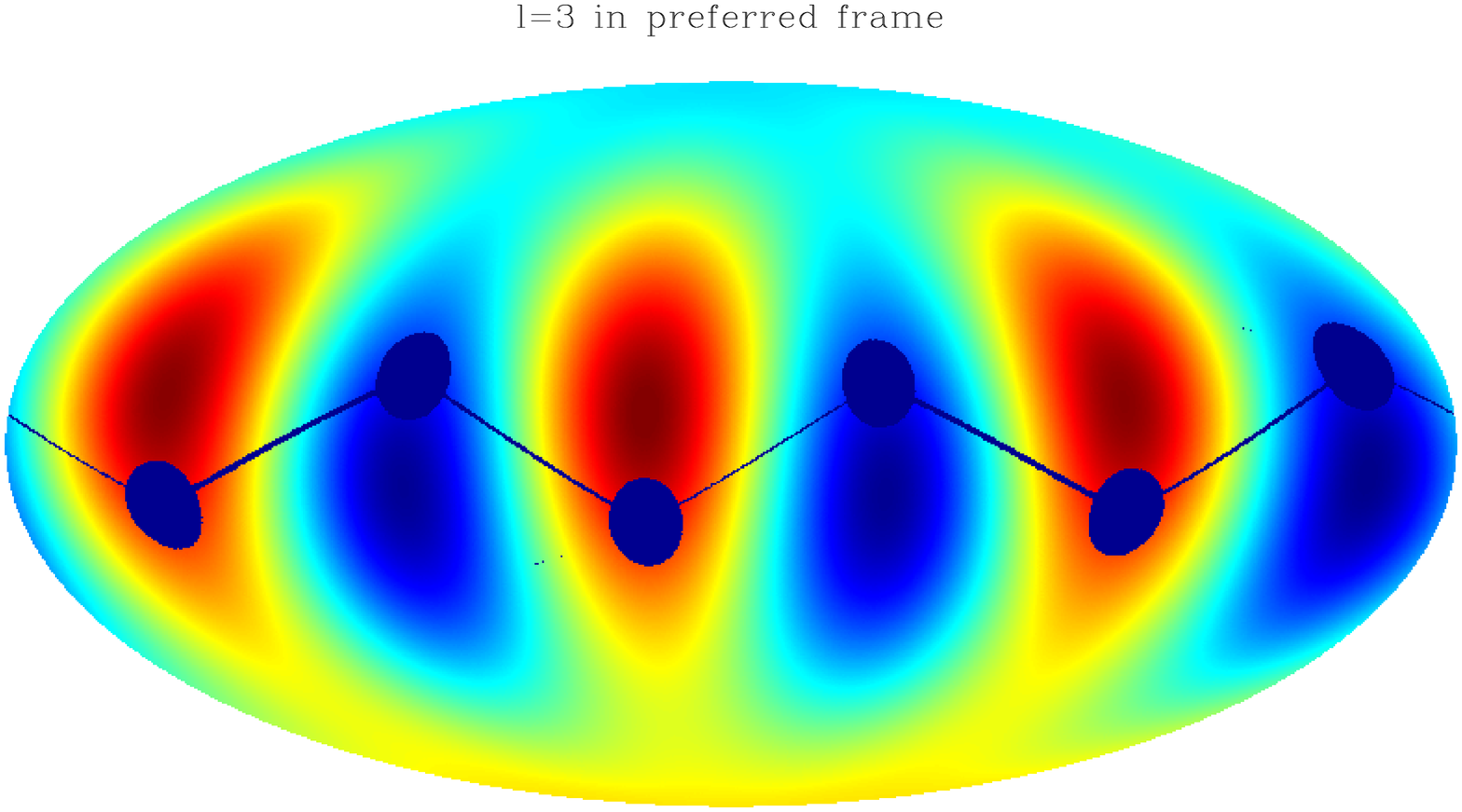,angle=90,width=9cm}}
\caption{The $\ell=5$ multipole for a cleaned map in galactic coordinates (top)
and aligned with $(b,l)\approx (50,-91)$ (middle.)
We have superposed the multipole vectors, and the
chain linking them. For comparison we plotted the $\ell =3$ multipole 
in its preferred frame.}
\label{fig1}
\end{figure}

It was suggested in~\cite{teg} that a possible statistic 
for anisotropy is 
\be
T_\ell=\max_{\bf n}\sum_m m^2 |a_{\ell m}|^2 
\ee
the average of $L_z^2$ maximized by an appropriate choice of
$z$-axis; the outcome of this statistic is twofold: the value $T_\ell$ and 
the axis ${\bf n}_\ell$. This statistic favours high $m$s and so it works well
in searches for planarity (i.e. $m=\pm \ell$). However the emergence
of a preferred axis could come about from undue concentration of power
in {\it any} $m$, a remark that is the main point of this letter.

We consider the alternative statistic:
\be
r_\ell=\max_{m{\bf n}} \frac{C_{\ell m}}
{(2\ell +1)C_\ell}
\ee
where $C_{\ell 0}=|a_{\ell 0}|^2$ and $C_{\ell m}=2|a_{\ell m}|^2$ 
for $m>0$ (notice that 2 modes contribute for $m\neq 0$).
This statistic provides three basic quantities:
the direction ${\bf n}_\ell$, the ``shape'' 
$m_\ell$, and the ratio $r_\ell$
of power absorbed by multipole $m_\ell$ in direction ${\bf n}_\ell$.
The statistic is never ambiguous except for planar $\ell=2$ 
configurations, which may be interpreted as $m=2$ or $m=1$
modes with axis ${\bf n}_2$ rotated by $90^\circ$. As
a convention we adopt the $(\ell=2,m=2)$ rendition for this
configuration.
In addition to the shape, direction, and strength of a given
multipole we may also define inter-$\ell$ quantities, such
as the angle between preferred axes for adjacent multipoles.

We have applied this method to a variety of maps: the cleaned and Wiener
filtered maps of~\cite{teg0}; the internal linear combination
maps~\cite{erik3} (see also~\cite{foregs}); the power equalized maps 
of~\cite{biel}. Taking cleaned and Wiener filtered maps,  
for $\ell=2,3$ 
our method promptly reproduces the results 
of~\cite{teg,copi}: we find planar multipoles aligned with
$(b,l)\approx(60,-100)$.  For higher $\ell$ we find some
surprising novelties. As an example, in  Fig.~\ref{fig1} we plot $\ell=5$
(top two panels). Galactic coordinates obscure the fact that 
the multipole is very approximately a pure $m=3$ mode aligned with 
$(b,l)=(50,-91)$. This is revealed by our statistic
and we have plotted the multipole when reoriented along its
preferred axis.
For comparison we plotted the $\ell=3$ mode aligned with its
preferred axis.

\begin{figure}
\centerline{\psfig{file=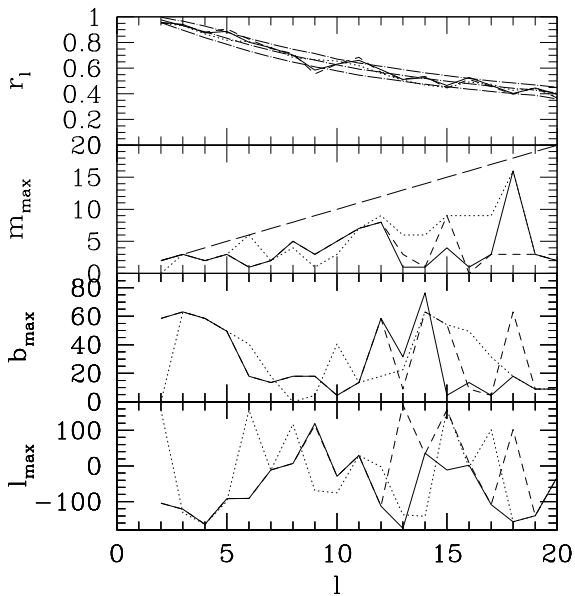,width=8cm}}
\caption{A plot of $m$-preference results up to $\ell=20$.
We seek the direction ${\mathbf n}_\ell$ (galactic coordinates
in the bottom two plots) where a given $m$ (second plot from the
top) receives the highest proportion of the power (with ratio
$r_{\ell}$ plotted in the top panel). We have also plotted the
Monte Carlo inferred variance for $r_\ell$. We plotted results
for cleaned (solid),  Wiener filtered (dash), and internal linear
combination (dotted) maps as described in text.}
\label{fig2}
\end{figure}

A complete application of this scheme to higher multipoles is 
shown in Fig.~\ref{fig2}. The multipole alignment
found for $\ell =2,...,5$ does not extend to higher
multipoles (even though there is a rough alignment of multipoles
$\ell=6,...,11$ around a strip on the $b\approx 15$ latitude).
However it is highly inconsistent with isotropy. 
We performed Monte Carlo simulations for Gaussian maps
with the best fit power spectrum, subject to the WMAP noise
and beam. We then considered the average value of all
possible angles between vectors ${\mathbf n}_\ell$ for $\ell=2,...,5$.
This is of the order of 20$^\circ$ for the data. We find that
only 5 out of 5000 simulations have an average value
smaller than this. Thus we may reject isotropy on these
scales at the 99.9\% confidence level.

Other apparent anomalous features, however, are not as 
statistically significant as they appear. The ratios
$r_{\ell}$, for instance,  are completely consistent with
Gaussianity, as the top panel of Fig.¬\ref{fig2} shows.
For example, for $\ell=3$ we find that in the preferred frame
94\% of the power is concentrated in the $m=3$ mode. This
was seen in~\cite{teg} as anomalous, at the 93\% confidence
level. Indeed this is true if we insist upon planarity. If 
on the contrary we allow for anisotropy due to {\it any} kind
of $m$-preference, then the majority of the simulations actually
find a $m$ and ${\mathbf n}$ with a {\it higher} concentration of 
power in a single $m$;
i.e. the observed $r_{\ell}$ is in fact {\it low}.

Likewise, the apparent preference for low $m$ multipoles 
(see Fig.~\ref{fig2}) is not exceptional, once a more rigorous 
evaluation of the feature is obtained from Monte Carlo simulations
(notice that the favoured $m$s are not uniformly distributed.)

However we do find another unusual feature with some statistical
significance. Phase correlations have been proposed as a discriminator
for localized non-Gaussian features in the sky~\cite{phasescoles}. 
In our formalism we may define $\ell$ azimuthal phases by examining
the phases of the $a_{\ell m}$, for $m>0$, as measured in the preferred
frame. Most important will be the phase of $a_{\ell m}$ for the $m$
that receives most of the power in this frame. This may be used
to fix the $x$ and $y$ axes of the preferred frame; because there
is an overall sign ambiguity between different $\ell$ and $m$ modes,
phases separated by $\pi$ should be identified. Inter-$\ell$ features 
may then be studied using the full set of Euler angles relating the 
preferred Cartesian frames for different multipoles.

We find a very close alignment of the phases for the $\ell=3$ and 
$\ell=5$ modes. This should be blatant from Fig.~\ref{fig1}, where
we see that the longitudinal stripes of the two modes align very 
strongly. Comparison with Monte-Carlo simulations reveals that this
feature is unusual to $94.5\%$. Other phases don't correlate significantly.

This concludes the presentation of our basic results, which may be summarised
as follows. The planar shape of the low $\ell$ multipoles is not statistically
significant, once we allow the simulations to seek an axis maximizing
power in a single, but {\it general} $m$.
However the alignment of the preferred axis 
is indeed very significant and extends up to $\ell=5$. An additional
azimuthal phase correlation brings to 99.995\% the confidence level
for rejecting Gaussianity and statistical isotropy.

We now examine the robustness of these results. Firstly they may
be confirmed using an  alternative technique:
the Maxwell multipole vectors. These were first introduced by 
Maxwell~\cite{maxwell} over a century ago and have found a recent revival 
within the CMB community~\cite{copi,den1,us1,den2,weeks1,weeks2,lach}.
They encode the $2\ell$ degrees of freedom contained 
in a multipole beside the power spectrum (see~\cite{us1,us2,den2} and
\cite{ral} for
how these split into invariants and ``eigenvectors'').
Multipole vectors may potentially pick up $m$-preference since for pure
$Y_{\ell m}$ configurations $\ell -m$ multipole vectors
align with the preferred direction ${\mathbf n}$, with the
remaining $m$ spreading at equal angles in the orthogonal plane --
we call this a ``handle and disc'' structure, and it provides
a more visual approach to detecting $m$-preference.

We consider these headless vectors in terms of pairs of 
points on the unit sphere. A handle is made through a
clustering of points; a disc through a set of points 
tracing out an equator on the sphere. 
We look for these patterns by searching for the shortest routes 
between the points. 
The method does not provide a thorough search 
for m-preference, but can be used as a more visual tool to confirm and explore 
already noted features. 
In Fig.~\ref{fig1} we show how the method 
fares with $\ell=3$ where we find a single disk, and $\ell=5$
where we find a nearly exactly planar disk of 3 vectors and a handle
of 2. More generally we confirm all the anomalous features
described above. Where the two methods disagree
there is no hint for $m$-preference.

How dependent are these results on the chosen data-set,
i.e. on the concrete rendition of the large-scale anisotropies 
after the galactic emissions have been removed? We find that 
most publicly available maps agree on the features outlined above and only
differ significantly beyond $\ell=12$. The cleaned and Wiener filtered maps 
of~\cite{teg0} in fact agree almost exactly up to  $\ell=12$
(see Fig.~\ref{fig2}). 
Some features below $\ell=12$, however, are not as robust when
we consider other maps. The quadrupole features, for instance,
may be easily erased in the renditions of~\cite{erik3,biel} 
(see Fig.~\ref{fig2}). The planarity of the $\ell=6$ mode is present
in some renditions (and is detected by our statistic) but not in 
others. Thus we should regard these features as more fickle.

\begin{figure}
\centerline{\psfig{file=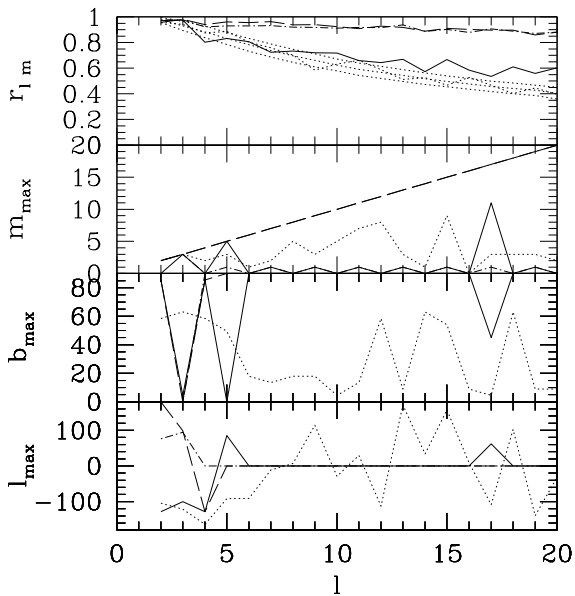,width=8cm}}
\caption{$m$-preference results up to $\ell=20$ for the V-band
of galactic templates. We have included free-free (solid) 
syncroton (dot-dash) and dust (long-dash) maps; for reference 
we have reproduced results for the cleaned maps (dotted line).}
\label{fig3}
\end{figure}

This leaves us with the obvious question: 
what could cause the strange directionality of the large angle
CMB fluctuations? Most obviously they could be due to systematic
errors or galactic foregrounds. We subjected galactic templates~\cite{foregs}
for  synchrotron, free-free and  dust emissions to this analysis
(Fig.~\ref{fig3}).
Unsurprisingly we found a strong non-Gaussian signal; however 
nothing correlated with the direction or signal
detected in the CMB maps. Specifically we found preference for
$m=0$ modes for even $\ell$ and for $m=1$ for odd $\ell$. This
preference is invariably found at $b=90$, i.e. with perfect alignment 
with the galaxy axis. The synchrotron and dust maps have very similar
morphologies; the free-free emission is quite different and leads
to more Gaussian $r_\ell$ ratios. None of this correlates with the
features claimed for the CMB, even when the templates are mixed
with a Gaussian component.

We also considered the simulated noise maps described 
in~\cite{systs}. These include all known systematic effects.
We found that {\it none} of the available 110 simulated maps
displayed any of the effects discussed in this paper.

Should the observed preferred axis be real our remarks may be crucial 
in identifying the culprit theory. We found that 
although there is a preferred direction for low multipoles
(and to some extent a preferred Cartesian axis)
this does not pick a specific $m$. Hence we don't need a model
favouring specific shapes (e.g. planarity); merely a model
with a preferred axis. If the low $\ell$ fluctuations are 
due to the gravitational potential on the last scattering surface
we can go further. The process may be described in terms
of the potential Fourier modes $\Phi({\mathbf k,\eta_{ls}})$; each of these
modes is reflected in the CMB in a pattern with exact $m=0$ shape,
with ${\mathbf n}$ aligned with
${\hat{\mathbf k}}$. One must superpose several such modes
to obtain a given $\ell$, specifically:
\be
a_{\ell m}=A\sum\Phi({\mathbf k,\eta_{ls}})
i^\ell j_\ell(k\Delta\eta_{ls})Y^\star_{\ell m}
({\hat{\mathbf k}})
\ee
Superposing a large number of modes leads to no preferred
direction or $m$ preference. Should the number of modes
be limited to a lattice, though, a preferred shape and/or
preferred axis will emerge. For a cubic lattice the low $\ell$ may become 
superpositions of mainly three modes $a_{\ell m}=A_1\delta_{m0}+A_2
Y^\star_{\ell m}({\pi\over 2},0) + A_3 Y^\star_{\ell m}({\pi\over 2},
{\pi\over 2})$ and a solution in terms of $A_i$ may now be found 
for a given observed $m$-preference. 

But it could also be that we live in a slab space, where there is a large
number of modes in all but one direction. This will erase any preference
for a specific $m$, while keeping the preferred direction. The 
choice between the two possibilities hinges crucially on the phase
correlation found for $\ell=3$ and $\ell=5$, with the implication
that there may be a preferred frame (rather than just a preferred
axis). We find this feature the most tantalising aspect of 
our analysis.

{\bf Acknowledgements} We thank Kris Gorski, Andrew Jaffe,
Jo\~ao Medeiros and Max Tgemark 
for helpful comments. Some of the results in this paper 
were derived using the HEALPix package (\cite{healp}), and calculations
were performed on COSMOS, the UK cosmology supercomputer facility.

\label{lastpage}

\end{document}